\documentclass[12pt]{article}
\usepackage{amsmath}
\usepackage{hyperref}
\usepackage{cite}
\usepackage{relsize}
\usepackage{amssymb}
\usepackage{graphics}
\usepackage{epsfig}
\usepackage{epstopdf}
\usepackage{verbatim}
\usepackage{color}
\usepackage{subcaption}
\usepackage{multirow}
\usepackage{textcomp}
\usepackage{eqnarray}
\usepackage{float}
\usepackage{mathtools}
\usepackage{pifont}
\usepackage{gensymb}
\usepackage{mathrsfs} 
\usepackage{array}
\usepackage{xcolor}
\newcolumntype{P}[1]{>{\centering\arraybackslash}p{#1}}
\usepackage[toc,page]{appendix}
\usepackage{authblk}
\usepackage{enumitem}
\usepackage{ragged2e}
\usepackage{dirtytalk}

\begin{document}

\title{\textbf{Predictive one-zero with vanishing sub-trace texture in neutrino mass matrix in light of dark matter and neutrinoless double beta decay}}
\author[1]{Ankush\thanks{ ankush.bbau@gmail.com}}
\author[2]{ Sangeeta Dey\thanks{ 
sangeetarimpi39@gmail.com}}
\author[3]{ Rishu Verma\thanks{ rishuvrm274@gmail.com}}
\author[4]{ Manoj Kumar\thanks{man1935kmr@gmail.com}}
\author[5]{B. C. Chauhan\thanks{ bcawake@hpcu.ac.in}}
\author[6]{Mahadev Patgiri\thanks{ mahadevpatgiri@cottonuniversity.ac.in}}

\affil[1]{School of Engineering and Technology\\CGC University, Mohali-140307, Punjab, India.}
\affil[2,6]{Department of Phyiscs\\Cotton University, Guwahati-781001, Assam, India.}
\affil[3,5]{Department of Physics and Astronomical Science\\ Central University of Himachal Pradesh, Dharamshala-176215, India.}
\affil[4]{Department of Physics, Akal University, Talwandi Sabo, Bathinda-151302, India. }

\maketitle
\begin{abstract}

\noindent In this work, we investigate a predictive class of neutrino mass matrices characterized by one texture zero and one vanishing sub-trace within the framework of the scotogenic model, wherein neutrino masses, dark matter, and neutrinoless double beta decay are intrinsically correlated. We analyze twelve viable texture structures---namely $B_{1,4,5}$, $C_{1,2,\ldots,5}$, $D_{4,5}$, and $F_{5,6}$---and examine their implications for the effective Majorana mass $(|M_{ee}|)$ governing neutrinoless double beta decay $(0\nu\beta\beta)$. Remarkably, all non-zero entries of the neutrino mass matrix can be parametrized in terms of this effective Majorana mass, establishing a direct theoretical link between low-energy observables and high-scale parameters of the model.
 Among the twelve textures, eleven predict dark matter masses of order TeV and yield correlated bounds on $|M_{ee}|$---making them testable in current and forthcoming $0\nu\beta\beta$ experiments---while the textures $D_4$ and $F_{5,6}$ exhibit comparatively weaker correlations. In contrast, the texture $C_5$ is excluded due to its requirement of unrealistically large Yukawa couplings and its inability to realize dark matter in the TeV regime. Our analysis thus identifies a subset of predictive neutrino mass textures that consistently relate dark matter phenomenology and neutrinoless double beta decay observables within the scotogenic paradigm.
\end{abstract}

\section{Introduction}
\label{sec:Introduction}

Although the Standard Model (SM) offers a robust framework to explain the fundamentals of particle interactions and unifies key forces, which makes it one of the most successful theories in physics, it still does not have a clear explanation for the non-zero neutrino mass. Therefore, one of the primary objective is to comprehend the origin of neutrino mass for which a number of beyond SM scenarios have been proposed in the literature, which are mainly of two types (a) by using the conventional tree level seesaw models \cite{Minkowski:1977sc,Mohapatra:1979ia,TY,Glashow:1979nm,Gell11} (b) using radiative mechanism. In particular, radiative seesaw models are attractive as they may account for the neutrino mass scales through suppression resulting from one loop \cite{Ma1,Zee:1980ai,Ma:2001mr,Kubo:2006yx,Hambye:2006zn,Farzan:2009ji}, two loop \cite{Ma:2007yx,Ma:2007gq,Kajiyama:2013zla,Aoki:2013gzs,Kajiyama:2013rla}, three loop and the higher loop levels \cite{Krauss:2002px,Aoki:2008av,Gustafsson:2012vj,Ahriche:2014cda,Ahriche:2014oda,Nomura:2016seu} and comparatively lighter mass of the mediator particle, which can be seen at current collider experiment. Dark matter and its characteristics represent yet another significant and unanswered puzzle in particle physics and cosmology. In many contexts, cosmological and astrophysical observations confirm that the current universe is made up of dark matter, an enigmatic, non-baryonic, non-luminous substance.  Therefore the fundamental scotogenic model which is proposed by E. Ma \cite{Ma1} is one of the most attractive radiative models at one-loop level, as it unifies the generation of neutrino mass with the physics of dark matter. The scotogenic model expands the scalar field content of the standard model by three fermion singlets and a scalar doublet, where $Z_{2}$ flavor symmetry is also augmented. The main purpose of $Z_2$ flavor symmetry is to stabilize the viable dark matter candidate, provided it is electrically neutral and it inhibits the scalar doublet from acquiring a non-zero vacuum expectation value (VEV), hence preventing the generation of neutrino mass at tree level. This makes the standard model (SM) neutrino masses scotogenic, meaning they emerge solely from radiative corrections within the dark sector. Thereby the small masses of the standard model neutrinos can be easily explained by the scotogenic model \cite{Ma1}.\\
 Further, the origin of neutrino masses and mixing have been be investigated through phenomenological approaches that impose specific structural constraints on the neutrino mass matrix, such as texture zeros (one or two entries in neutrino mass matrix are zero) \cite{t10,t20,t30,t40}, hybrid textures (one zero and an equality among elements) \cite{ht1,ht2,ht3,ht4,ht5} etc., independent of the details of the underlying theory. These ansatz offer enhanced predictability by reducing the number of free parameters in the neutrino mass matrix. In the work \cite{Ankush:2021opd}, one of the authors, have discussed hybrid textures in the scotogenic model framework leading to predictive scenarios in light of dark matter and neutrinoless double beta decay. Also, determining the nature of neutrinos i.e., whether they are Dirac or Majorana particles, is another important challenge in the field of neutrino physics. Consequently, from the phenomenological perspective, it is quite significant to investigate the neutrinoless double beta decay $(0\nu\beta\beta)$, as this process, if experimentally confirmed in the future, may reveal the Majorana nature of the neutrinos.  \\
In this paper, we are mainly motivated to address the above mentioned problems in the framework of Majorana neutrino mass matrix, by constraining the matrix with the condition of one zero element and one vanishing sub-trace. The one zero textures of the neutrino mass matrix with the condition of vanishing sub-trace has been discussed by the authors in their work \cite{Dey:2025yjy}. Here, we mainly focus on carrying our work with the twelve textures which are allowed for normal mass hierarchy (NH) at $3\sigma$ neutrino oscillation data. Interestingly, all the elements of the neutrino mass matrix can be expressed in terms of $|M_{ee}|$, except for the one zero elements.\\ The paper is organized as follows: In Sec.2 we discuss the scotogenic model and relic density of dark matter, followed by the dark matter and neutrinoless double beta decay in a one-zero texture framework with vanishing sub-trace in Sec.3. In Sec.4 we have shown the numerical analysis and discussion. Finally, the conclusions are summarized in Sec.5.
\section{Scotogenic Model and Relic Density of Dark Matter} We extend the Standard Model (SM) by introducing three right-handed singlet fermions \( N_k \) (for \( k = 1, 2, 3 \)), which are singlets under \( SU(2)_L \), along with a new scalar doublet \( (\eta^+, \eta^0) \) that transforms as a doublet under \( SU(2)_L \). Both the newly added fermions and the scalar doublet are odd under an exact \( Z_2 \) symmetry, as proposed in \cite{Ma1}. The transformation properties of the particles under the symmetry group \( SU(2)_L \times U(1)_Y \times Z_2 \) are summarized in Table \ref{particle content}.

\begin{table}[h]
    \centering
   \begin{tabular}{ |p{3cm}|p{1.5cm}|p{1.5cm}|p{1.5cm}|}
\hline
Particle Content & \multicolumn{3}{c|}{Charges}\\
\cline{2-4}
&  $SU(2)_L$ & $U(1)_Y$ & $Z_2$ \\
\hline
$L_{\alpha}=(\nu_{\alpha},l_{\alpha})$ & 2 &-1/2& + \\ \hline
$l^{C}_{\alpha}$ & 1   & 1 & +\\ \hline
$\phi=(\phi^{+},\phi^{0})$ &2 & -1/2&+ \\ \hline
$\eta=(\eta^{+},\eta^{0})$    &2 & 1/2 &-\\ \hline
$N_{k}$ & 1 & 0 &- \\

\hline
\end{tabular}
    \caption{The particle content of the model under $SU(2)_{L}\times U(1)_{Y} \times Z_{2}$.}
    \label{particle content}
\end{table}

Here, $\alpha=e,\mu,\tau$,  $(\nu_{\alpha},l_{\alpha})$  are left-handed lepton doublets and $(\phi^{+},\phi^{0})$ is Higgs doublet.\\
The Lagrangian of the model, including the relevant Yukawa interactions and mass terms, is expressed as follows:
\begin{equation}
    \mathcal{L} \supset h_{\alpha k}(\bar{\nu}_{\alpha L}\eta^{0}-\bar{l}_{\alpha L} \eta^{+})N_{k} +\frac{1}{2}M_{k}\bar{N}_{k}N^{C}_{k} +h.c.,
    \label{2.2}
\end{equation}
and the corresponding scalar potential interaction terms of interest are given by:
\begin{equation}
    V\supset \frac{1}{2}\lambda (\phi^{\dagger}\eta)^{2} +h.c.,
    \label{2.3}
\end{equation}
here, $\lambda$ denotes the quartic coupling. Due to the presence of an exact $Z_2$ symmetry, neutrino masses are forbidden at the tree level. As a result, the neutrino masses are generated radiatively at the one-loop level. The general element of the neutrino mass matrix, $M_{\alpha \beta}$ is expressed as

\begin{equation}
    M_{\alpha \beta} = \sum _{k=1}^{3} h_{\alpha k}h_{\beta k}\Lambda_{k},
    \label{2.4}
\end{equation}
where
\begin{equation}
    \Lambda_{k}=\frac{\lambda v^{2}}{16 \pi^{2}}\frac{M_{k}}{m_{0}^{2}-M_{k}^{2}}\left(1-\frac{M_{k}^{2}}{m_{0}^{2}-M_{k}}\ln\frac{m_{0}^{2}}{M_{k}^{2}}\right),
    \label{2.5}
\end{equation}
\begin{equation}
    m_{0}^{2}=\frac{1}{2}(m_{R}^{2}+m_{I}^{2}).
    \label{2.6}
\end{equation}
Here, $\beta = (e, \mu, \tau)$ represents the lepton flavors, and $v = 246~\text{GeV}$ denotes the vacuum expectation value (vev) of the Higgs field. The parameters $m_{R}$ and $m_{I}$ correspond to the masses of $\sqrt{2}~\text{Re}[\eta^{0}]$ and $\sqrt{2}~\text{Im}[\eta^{0}]$, respectively, while $M_{k}~(k=1,2,3)$ are the masses of the right-handed neutrinos.\\
\noindent At the level of one loop, lepton flavor violating (LFV) processes like $\mu\to e\gamma$ are induced. Consequently, the branching ratio for the $\mu \to e \gamma$ process is expressed as \cite{Kubo:2006yx,Ma:2001mr}
\begin{equation}
    Br(\mu\to e\gamma)=\frac{3 \alpha_{em}}{64 \pi(G_{F}m_{0}^{2})^{2}}\left|\sum_{k=1}^{3}h_{\mu k}h_{e k}^{*}F\left(\frac{M_{k}}{m_{0}}\right)\right|^{2}.
    \label{2.7}
\end{equation}
Here, $\alpha_{em}$ denotes the fine-structure constant associated with electromagnetic interactions, $G_{F}$ is the Fermi coupling constant, and the function $F(r)$ is defined as

\begin{equation}
F(r)=\frac{1-6r^{2}+3r^{4}+2r^{6}-6r^{4}  \ln r^{2}}{6(1-r^{2})^{4}} , \hspace{0.2cm} r\equiv \frac{M_{k}}{m_{0}}.  
\label{2.8}
\end{equation}

\noindent It is noteworthy that this model simultaneously accounts for both dark matter (DM) and neutrino mass generation. The lightest of the $Z_2$-odd fermions, $N_k$, is stable and hence constitutes a plausible dark matter candidate. Incorporating coannihilation effects, the model may concurrently account for the observed relic abundance of cold dark matter and comply with the experimental constraints on the branching ratio of the lepton flavour violating (LFV) process $\mu \to e \gamma$. In this analysis, we assume that the mass of the dark matter particle, $N_1$ is nearly degenerate with that of the next lightest singlet fermion $N_2$, and the mass hierarchy among the right-handed neutrinos is given by $M_1 \leq M_2 < M_3$ \cite{Griest}. The product of the coannihilation cross-section and the relative velocity $v_r$ of the annihilating particles is given by \cite{Suematsu:2009ww}

\begin{equation}
    \sigma_{ij}|v_{r}|=a_{ij}+b_{ij}v_{r}^{2},
    \label{2.9}
\end{equation}
where

\begin{align}
\label{2.10}
\left.
 \begin{array}{lll}
     a_{ij}=\frac{1}{8\pi} \frac{M_{1}^2}{(M_{1}^2+m_{0}^{2})^{2}}\sum_{\alpha, \beta}(h_{\alpha i}h_{\beta j}-h_{\alpha j}h_{\beta i})^{2},\\
    b_{ij}=\frac{m_{0}^{4}-3m_{0}^{2}M_{1}^{2}-M_{1}^{4}}{3(M_{1}^{2}+m_{0}^{2})^{2}} a_{ij}+\frac{1}{12 \pi}\frac{M_{1}^{2}(M_{1}^{4}+m_{0}^{4})}{(M_{1}^{2}+m_{0}^{2})^{4}}\sum_{\alpha,\beta} h_{\alpha i}h_{\alpha j}h_{\beta i}h_{\beta j},
 \end{array}
 \right\}.
 \end{align}
In Eqn.(\ref{2.9}), $\sigma_{ij}$ $(i,j = 1,2)$ denotes the annihilation cross-section for the process $N_i N_j \to X \bar{X}$. The parameter $dM = (M_2 - M_1)/M_1$ represents the mass splitting ratio between the nearly degenerate singlet fermions, and $x = M_1/T$ corresponds to the ratio of the dark matter mass to the temperature $T$. If $g_{1}$, $g_{2}$ are the number of degrees of  freedom of singlet fermions $N_{1}$ and $N_{2}$, respectively, the effective cross section is given by
\begin{equation}
\label{a2}
    \sigma_{eff}=\frac{g_{1}^{2}}{g_{eff}^{2}}\sigma_{11}+\frac{2g_{1}g_{2}}{g_{eff}^{2}}\sigma_{12}(1+dM)^{3/2}\exp(-xdM)+\frac{g_{2}^{2}}{g_{eff}^{2}}\sigma_{22}(1+dM )^{3}\exp(-2 xdM),
\end{equation}
\begin{equation}
\label{a3}
    g_{eff}=g_{1}+g_{2}(1+dM)^{3/2} \exp(-xdM),
\end{equation}
with $dM \simeq 0$ ($N_{1}$ is almost degenerate with $N_{2}$) and using Eqns.(\ref{2.9}) and (\ref{a3}) in Eqn.(\ref{a2}) we get\\
The thermal average cross section is
\begin{equation}
    <\sigma_{eff}|v_{r}|> = a_{eff}+6 b_{eff}/x,
    \label{2.15}
\end{equation}
where 
\begin{align}
\label{2.14}
    \left.
    \begin{array}{cc}
      a_{eff}=\frac{a_{11}}{4}+\frac{a_{12}}{2}+\frac{a_{22}}{4},\\
          b_{eff}=\frac{b_{11}}{4}+\frac{b_{12}}{2}+\frac{b_{22}}{4},
\end{array}
\right\}.
\end{align}

\noindent Now, the relic density of cold dark matter is given by

\begin{equation}
\label{2.16}
    \Omega h^{2} =\frac{1.07\times 10^{9} \text{GeV}^{-1} x_{f}}{(a_{eff}+3 b_{eff}/x_{f}) g_{*}^{1/2} m_{Pl}},
\end{equation}
where $m_{Pl}=1.22\times10^{19}$GeV , $g_{*}=106.75$
and
\begin{equation}
 x_{f}=\ln \frac{0.038 g_{eff} m_{Pl} M_{1}<\sigma_{eff}|v_{r}|> }{g_{*}^{1/2}x_{f}^{1/2} }.   \label{2.17}
\end{equation}
Also $x_{f}= \frac{M_1}{T_f}\approx 25$, $T_{f}$ is the freeze-out temperature\cite{Kolb:1990vq}.

\section{Dark Matter and Neutrinoless Double Beta Decay in a One-Zero Texture Framework with Vanishing Sub-trace } 
After establishing the basic framework of the scotogenic model and outlining the methodology for calculating the relic density of dark matter, we now proceed to analyze the neutrino mass matrix.
The general form of the neutrino mass matrix can be constructed as  
\\
\begin{equation}
M_{\nu}= U diag (m_{1},m_{2}e^{i\alpha_2},m_{3}e^{i\alpha_3}) U^{T}\equiv M_{\alpha\beta},
\label{3.1}
\end{equation}
 \\
where $U=f(\theta_{12}, \theta_{23}, \theta_{13},\delta)$, denotes the Pontecorvo-Maki-Nakagawa-Sakata (PMNS) mixing matrix, $m_i,\hspace{0.1cm} i=1,2,3$ represents the three neutrino mass eigenvalues, ($\theta_{12},\theta_{23},\theta_{13}$) are mixing angles, $\delta$ is the Dirac CP phase, $\alpha_{2,3}$ are the Majorana phases and $(\alpha,\beta=e,\mu,\tau)$. 
Therefore, elements of the neutrino mass matrix can be written as
\begin{equation}
     M_{\alpha\beta }=f(m_1,m_2,m_3,\theta_{12},\theta_{23},\theta_{13},\alpha_2,\alpha_3,\delta).
   \label{3.2}
 \end{equation}

 \noindent
However, the non-zero elements of the neutrino mass matrix can be represented as being proportional to $M_{ee}$, the effective Majorana mass\cite{Kitabayashi:2015jdj,Kitabayashi:2015tka,Ankush:2021opd,Ankush:2023pax}.
\begin{equation}
M_{\alpha\beta}=f_{\alpha\beta}(\theta_{12},\theta_{23},\theta_{13},\delta) M_{ee},
\label{3.4}
\end{equation} 
where ($\alpha,\beta=e,\mu,\tau$). \\
From Eqn.(\ref{3.4}), it can be seen that any element of the neutrino mass matrix can be expressed as proportional to $M_{ee}$, i.e., the effective Majorana neutrino mass. This is done in order to study the correlation between the dark matter and neutrinoless double beta decay amplitude. Therefore, we can express the elements of the texture structures of the neutrino mass matrices constrained with the condition of one zero element and one vanishing sub-trace. The study of these texture structures are already being carried in the literature  where it is seen that out of thirty-six possible structures of neutrino mass matrices, only thirteen textures are phenomenologically allowed for Normal Hierarchy (NH) in light of the $3\sigma$ neutrino oscillation data \cite{Dey:2025yjy}. In this work we carry out the study of only thirteen possible textures shown in Table(\ref{tab3003}), which are consistent for the normal mass ordering.

\begin{table}[ht]
      \centering
\resizebox{17cm}{!}{\begin{tabular}{|c|c|c|c|c|} 
\hline
$A_1$ & $B_1$  & $B_4$ & $B_5$ & $C_1$  \\
\hline
$\begin{pmatrix}
0 & m_{e\mu} & m_{e\tau}\\
m_{e\mu} & m_{\mu\mu} & m_{\mu\tau}\\
m_{e\tau} & m_{\mu\tau} & -m_{\mu\mu}\\
\end{pmatrix}$ & $\begin{pmatrix}
m_{ee} & 0 & m_{e\tau}\\
0 & m_{\mu\mu} & m_{\mu\tau}\\
m_{e\tau} & m_{\mu\tau} & -m_{\mu\mu}\\
\end{pmatrix}$ & $\begin{pmatrix}
m_{ee} & 0 & m_{e\tau}\\
0 & m_{\mu\mu} & m_{\mu\tau}\\
m_{e\tau} & m_{\mu\tau} & -m_{ee}\\
\end{pmatrix}$ & $\begin{pmatrix}
m_{ee} & 0 & m_{e\tau}\\
0 & m_{\mu\mu} & -m_{ee}\\
m_{e\tau} & -m_{ee} & m_{\tau \tau}\\
\end{pmatrix}$ & $\begin{pmatrix}
m_{ee} & m_{e\mu} & 0\\
m_{e\mu} & m_{\mu\mu} & m_{\mu\tau}\\
0 & m_{\mu\tau} & -m_{\mu\mu}\\
\end{pmatrix}$ \\ 
\hline
$C_2$ & $C_3$  & $C_4$ & $C_5$ & $D_4$  \\
\hline
$\begin{pmatrix}
m_{ee} & m_{e\mu} & 0\\
m_{e\mu} & m_{\mu\mu} & m_{\mu\tau} \\
0 & m_{\mu\tau} & -m_{e\mu}\\
\end{pmatrix}$ & $\begin{pmatrix}
m_{ee} & m_{e\mu} & 0\\
m_{e\mu} & m_{\mu\mu} & -m_{e\mu}\\
0 & -m_{e \mu} & m_{\tau \tau}\\
\end{pmatrix}$ & $\begin{pmatrix}
m_{ee} & m_{e\mu} & 0\\
m_{e\mu} & m_{\mu\mu} & m_{\mu\tau}\\
0 & m_{\mu\tau} & -m_{ee}\\
\end{pmatrix}$ & $\begin{pmatrix}
m_{ee} & m_{e\mu} & 0\\
m_{e\mu} & m_{\mu\mu} & -m_{ee}\\
0 & -m_{ee} & m_{\tau \tau}\\
\end{pmatrix}$ & $\begin{pmatrix}
m_{ee} & m_{e\mu} & m_{e\tau}\\
m_{e\mu} & 0 & m_{\mu\tau}\\
m_{e\tau} & m_{\mu\tau} & -m_{ee}\\
\end{pmatrix}$ \\
\hline
$D_5$ & $F_5$  & $F_6$ & {} & {}  \\
\hline
$\begin{pmatrix}
m_{ee} & m_{e\mu} & m_{e\tau}\\
m_{e\mu} & 0 & -m_{ee}\\
m_{e\tau} & -m_{ee} & m_{\tau \tau}\\
\end{pmatrix}$ & $\begin{pmatrix}
m_{ee} & m_{e\mu} & m_{e\tau}\\
m_{e\mu} & m_{\mu\mu} & -m_{ee}\\
m_{e\tau} & -m_{ee} & 0\\
\end{pmatrix}$ & $\begin{pmatrix}
m_{ee} & m_{e\mu} & m_{e\tau}\\
m_{e\mu} & -m_{ee} & m_{\mu\tau}\\
m_{e\tau} & m_{\mu\tau} & 0\\
\end{pmatrix}$ & {} & {} \\
\hline
\end{tabular}}
\caption{Thirteen allowed textures for NH at $3\sigma$ C.L \cite{Dey:2025yjy}.}
\label{tab3003}
 \end{table}

\noindent  
As the $(1,1)$ element of texture $A_1$ is zero, therefore, this texture will not contribute to the study of $0\nu\beta\beta$-decay. Therefore, we will not consider this texture for our analysis. We shall continue our study with the remaining twelve textures which are allowed only for a constrained range of the Dirac CP phase, $\delta$, as shown in Table(\ref{tab303}).\\ 

\begin{table}[ht]
      \centering
\begin{tabular}{ |c|c|c| } 
\hline
Class & Textures  & Range of $\delta$ \\
\hline

{} & $B_1$ & $(270^{\degree}- 288^{\degree})$ \\ 

B & $B_4$ & $(342^{\degree}- 346^{\degree})$ \\ 

{} & $B_5$ & $(160^{\degree} - 250^{\degree})$ \\ 
\hline

{} & $C_1$ & $(250^{\degree}- 275^{\degree})$ \\ 

{} & $C_2$ & $(240^{\degree}- 258^{\degree})$ \\ 

C & $C_3$ & $(144^{\degree} - 350^{\degree})$ \\ 

{} & $C_4$ & $(220^{\degree} - 230^{\degree})$ \\

{} & $C_5$ & $(270^{\degree} - 350^{\degree})$ \\
\hline
D & $D_4$ & $(144^{\degree} - 350^{\degree})$ \\ 

{} & $D_5$ & $(220^{\degree} - 230^{\degree})$ \\
\hline
F & $F_5$ & $(144^{\degree} - 350^{\degree})$ \\ 

{} & $F_6$ & $(144^{\degree} - 350^{\degree})$ \\
\hline
\end{tabular}
\caption{Allowed Range of $\delta$ for all the twelve textures \cite{Dey:2025yjy}.}
\label{tab303}
 \end{table}
Further, the remaining twelve textures can be written in form of Eqn.(\ref{3.4}). In general, the elements of the neutrino mass matrix for each texture are written as

\begin{equation}
M_{\alpha\beta}=f^\mathscr{X}_{\alpha\beta}(\theta_{12},\theta_{23},\theta_{13},\delta) M_{ee},
\label{3.5}
\end{equation} 
where ($\alpha,\beta=e,\mu,\tau$) and $\mathscr{X}=B_{1,4,5}$, $C_{1,2,...5}$, $D_{4,5}$, $F_{5,6}$. \\

Since $(1,1)$ element of each matrix represents $M_{ee}$ itself, the coefficients $f_{ee}$ for all these flavor structures are unity. We can write all the textures, $B_{1,4,5}$, $C_{1,2,...5}$, $D_{4,5}$, $F_{5,6}$ in more simplified form as\\
\;
\;
\;
\;
\;
\;
\;

\begin{align}
\left.
 \begin{array}{ll}
B_1:\begin{pmatrix}
1 & 0 & f_{e\tau}^{B_1}\\
0 & f_{\mu\mu}^{B_1} & f_{\mu\tau}^{B_1}\\
f_{e\tau}^{B_1} & f_{\mu\tau}^{B_1} & -f_{\mu\mu}^{B_1}\\
\end{pmatrix}M_{ee},\hspace{1.8In}
 B_4:\begin{pmatrix}
f_{ee}^{B_4} & 0 & f_{e\tau}^{B_4}\\
0 & f_{\mu\mu}^{B_4} & f_{\mu\tau}^{B_4}\\
f_{e\tau}^{B_4} & f_{\mu\tau}^{B_4} & -f_{ee}^{B_4}\\
\end{pmatrix} M_{ee},\vspace{0.8cm} \\ 
 B_5:\begin{pmatrix}
1 & 0 & f_{e\tau}^{B_5}\\
0 & f_{\mu\mu}^{B_5} & -f_{ee}^{B_5}\\
f_{e\tau}^{B_5} & -f_{ee}^{B_5} & f_{\tau \tau}^{B_5}\\
\end{pmatrix} M_{ee} 
\end{array}
\right\}
\end{align}
\begin{align}
\left.
 \begin{array}{lll}
    C_1:\begin{pmatrix}
1 & f_{e\mu}^{C_1} & 0\\
f_{e\mu}^{C_1} & f_{\mu\mu}^{C_1} & f_{\mu\tau}^{C_1}\\
0 & f_{\mu\tau}^{C_1} & -f_{\mu\mu}^{C_1}\\
\end{pmatrix}M_{ee},\hspace{1.8In}
C_2:\begin{pmatrix}
1 & f_{e\mu}^{C_2} & 0\\
f_{e\mu}^{C_2} & f_{\mu\mu}^{C_2} & f_{\mu\tau}^{C_2}\\
0 & f_{\mu\tau}^{C_2} & -f_{e\mu}^{C_2}\\
\end{pmatrix}M_{ee},\vspace{0.8cm} \\
C_3:\begin{pmatrix}
1 & f_{e\mu}^{C_3} & 0\\
f_{e\mu}^{C_3} & f_{\mu\mu}^{C_3} & -f_{e\mu}^{C_3}\\
0 & -f_{e \mu}^{C_3} & f_{\tau \tau}^{C_3}\\
\end{pmatrix}M_{ee}\hspace{1.8In}
C_4:\begin{pmatrix}
1 & f_{e\mu}^{C_4} & 0\\
f_{e\mu}^{C_4} & f_{\mu\mu}^{C_4} & f_{\mu\tau}^{C_4}\\
0 & f_{\mu\tau}^{C_4} & -1\\
\end{pmatrix}M_{ee},\vspace{0.8cm} \\
C_5:\begin{pmatrix}
1 & f_{e\mu}^{C_5} & 0\\
f_{e\mu}^{C_5} & f_{\mu\mu}^{C_5} & -1\\
0 & -1 & f_{\tau \tau}^{C_5}\\
\end{pmatrix}M_{ee},
\hspace{3.6In}
\end{array}
\right\}
\end{align}

\begin{equation}
    D_4:\begin{pmatrix}
1 & f_{e\mu}^{D_4} & f_{e\tau}^{D_4}\\
f_{e\mu}^{D_4} & 0 & f_{\mu\tau}^{D_4}\\
m_{e\tau}^{D_4} & m_{\mu\tau}^{D_4} & -1\\
\end{pmatrix}M_{ee}, \hspace{1.8In}
D_5:\begin{pmatrix}
1 & f_{e\mu}^{D_5} & f_{e\tau}^{D_5}\\
f_{e\mu}^{D_5} & 0 & -1\\
f_{e\tau}^{D_5} & -1 & f_{\tau \tau}^{D_5}\\
\end{pmatrix}M_{ee}
\end{equation}
\begin{equation}
    F_5:\begin{pmatrix}
1 & f_{e\mu}^{f_5} & f_{e\tau}^{f_5}\\
f_{e\mu}^{f_5} & f_{\mu\mu}^{f_5} & -1\\
f_{e\tau}^{D_5} & -1 & 0\\
\end{pmatrix}M_{ee}, \hspace{1.8In}
F_6:\begin{pmatrix}
1 & f_{e\mu}^{f_6} & f_{e\tau}^{f_6}\\
f_{e\mu}^{f_6} & -1 & f_{\mu\tau}^{f_6}\\
f_{e\tau}^{f_6} & f_{\mu\tau}^{f_6} & 0\\
\end{pmatrix}M_{ee},
\end{equation}

\;
\;
\;
\;
\;
\;

where the coefficients $f_{\alpha \beta}^{\mathscr{X}}$  are given in Appendix. Furthermore, these coefficients $f_{\alpha \beta}^{\mathscr{X}}$ are computed by randomly varying the mixing angles and CP-violating phase within their experimentally allowed ranges, as specified in Table~\ref{tab33}~\cite{Esteban:2018azc}.

\;
\;
\;
\;
\;
\;
\;
\;

  \begin{table}
      \centering
\begin{tabular}{ |c|c|c|c| } 
\hline
Mixing angles & bfp $\pm$ 1$\sigma$ & 3$\sigma$ range \\
\hline
$\theta_{12}/^{o}$ & $33.82^{+0.78}_{-0.76}$ & $31.61\rightarrow36.27$ \\ 
\hline
$\theta_{23}/^{o}$ & $49.6^{+1.0}_{1.2}$ & $40.3\rightarrow52.4$ \\ 
\hline

$\theta_{13}/^{o}$ & $8.61^{+0.13}_{-0.13}$ & $8.22\rightarrow8.99$ \\ 
\hline

\end{tabular}
\caption{Global fit data of neutrino mixing angles \cite{Esteban:2018azc}.}
\label{tab33}
 \end{table}
  \begin{table}
      \centering
      \begin{tabular}{|c|c|}
      \hline
       Parameter    & Range \\ \hline
        $\lambda$   & ($3-4$)$\times 10^{-9}$\\
        \hline
        $h_{e1},h_{\mu2},h_{\tau3}$ &  $0-1.5$
        \\
        \hline
        $M_1$ & $\mathcal{O}(TeV)$\\
        \hline
      \end{tabular}
      \caption{Ranges of parameters used in the numerical analysis.}
      \label{tab3}
  \end{table}

\section{Numerical Analysis and Discussion}
\begin{figure}[H]
    \centering
    \includegraphics[width=1\linewidth]{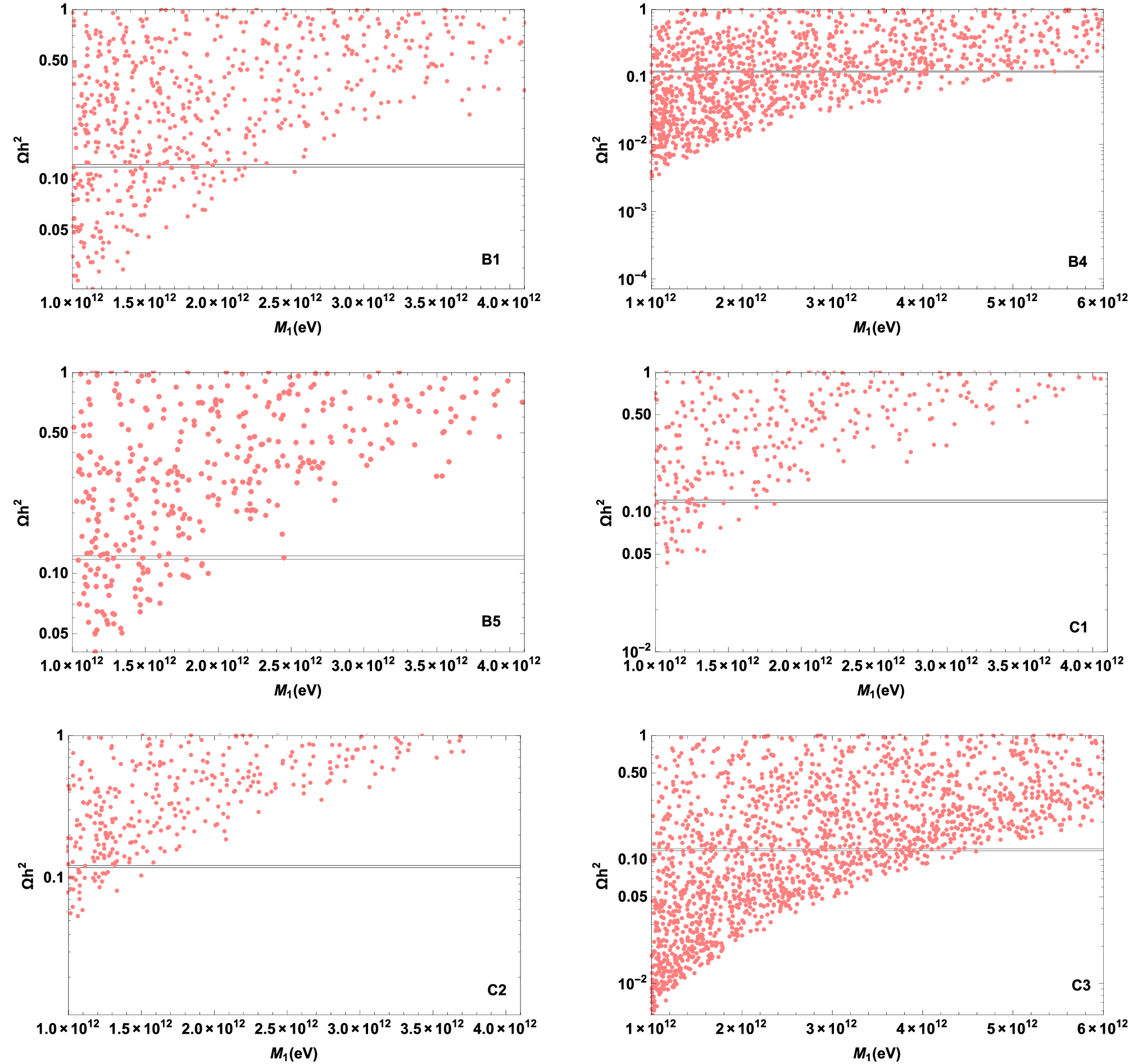}
    \caption{Correlation between Relic density of dark matter $(\Omega h^2)$ and dark matter mass $(M_1)$ for textures $B_{1,4,5}$, $C_{1,2,3}$. The horizontal lines indicating the measured relic density value, $\Omega h^2 = 0.1191 \pm 0.0022$ \cite{Planck}.}
    \label{M1}
\end{figure}

\begin{figure}[H]
    \centering
    \includegraphics[width=1\linewidth]{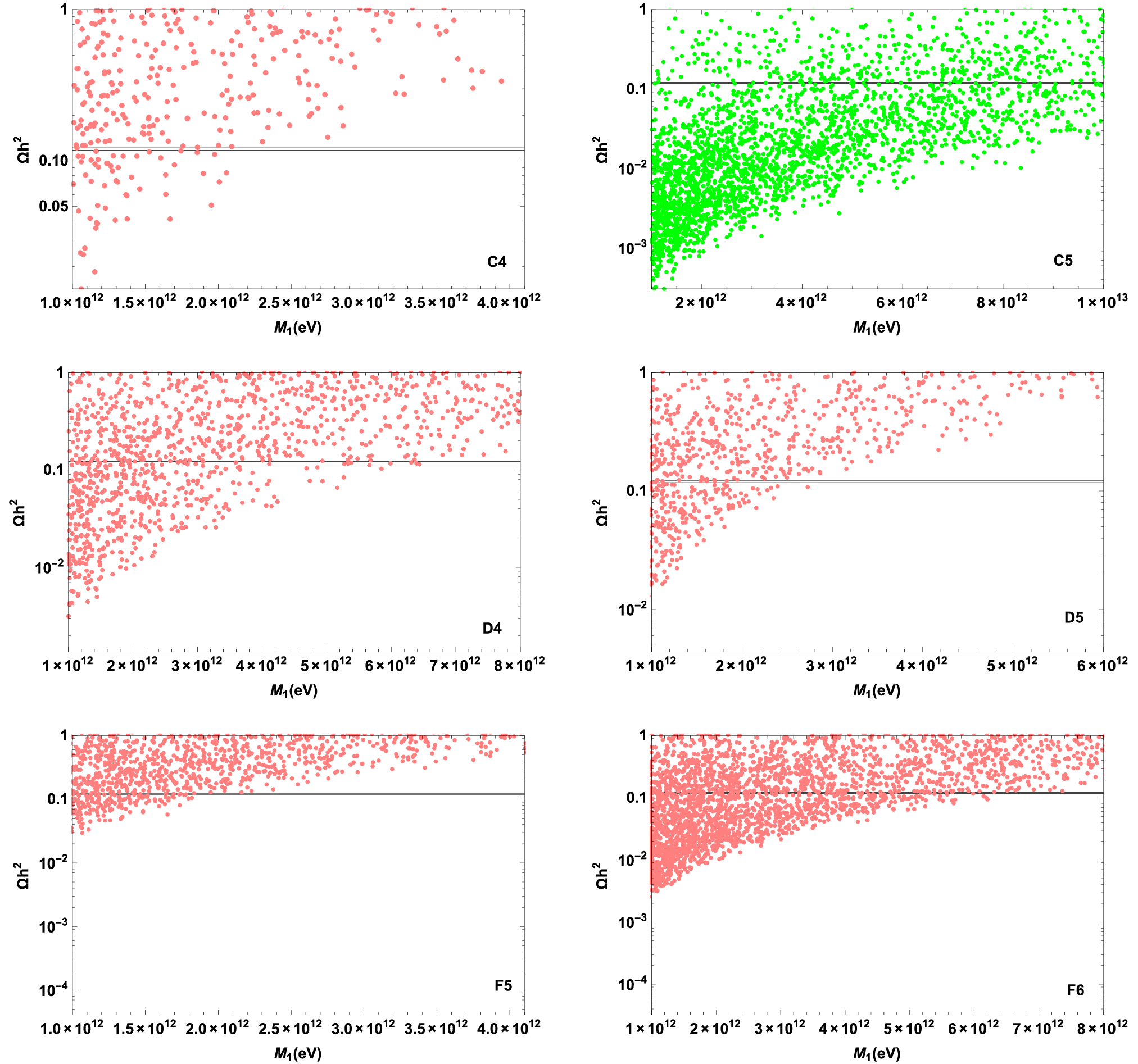}
    \caption{Correlation between Relic density of dark matter $(\Omega h^2)$ and dark matter mass $(M_1)$ for textures $C_4$,$D_{4,5}$, $F_{5,6}$. The horizontal lines indicating the measured relic density value, $\Omega h^2 = 0.1191 \pm 0.0022$ \cite{Planck}.}
    \label{M2}
\end{figure}

\begin{figure}[H]
    \centering
    \includegraphics[width=1\linewidth]{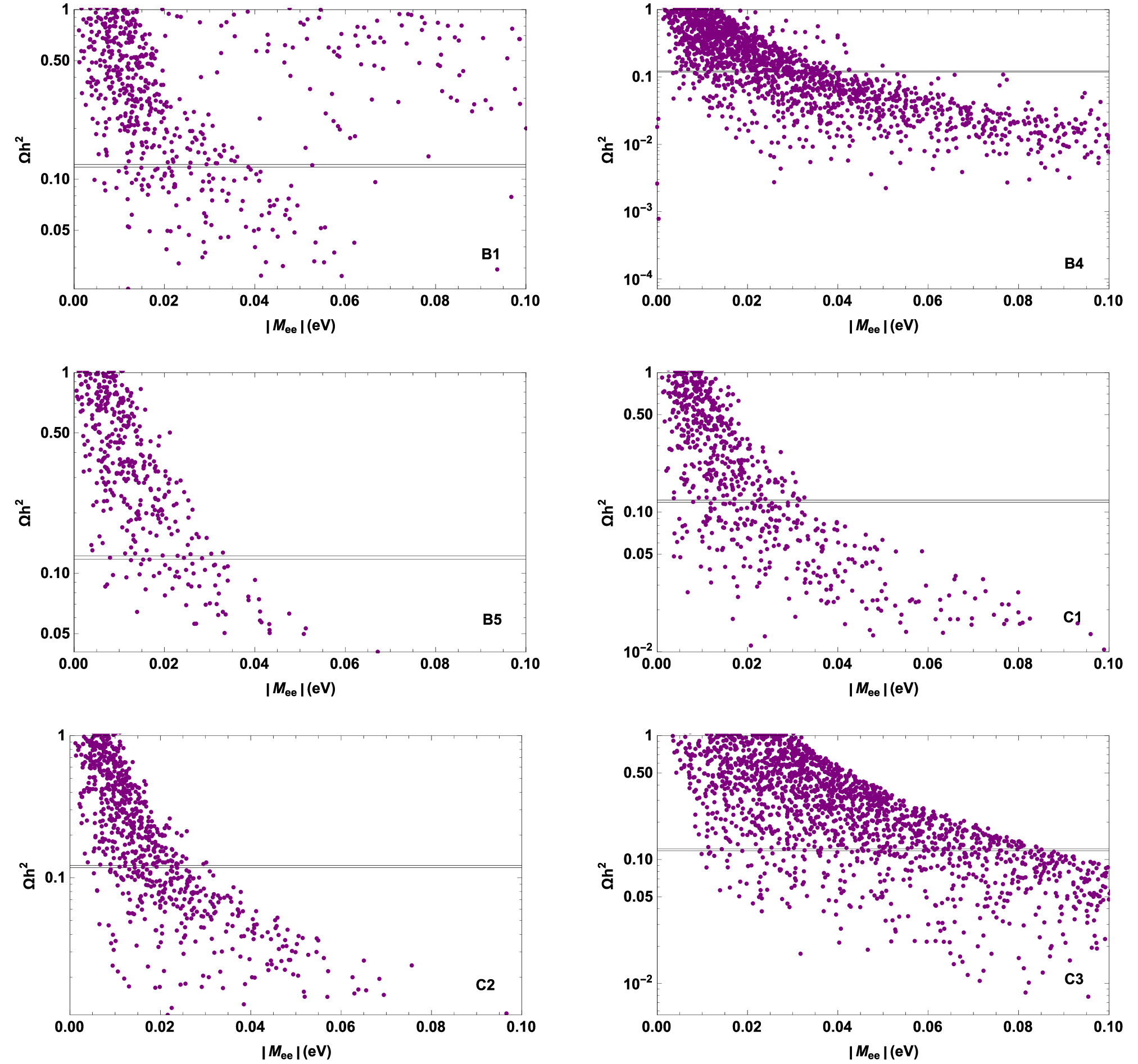}
    \caption{Correlation between Relic density of dark matter $(\Omega h^2)$ and effective Majorana mass $(|M_{ee}|)$ in neutrinoless double beta decay for textures $B_{1,4,5}$, $C_{1,2,3}$. The horizontal lines indicating the measured relic density value, $\Omega h^2 = 0.1191 \pm 0.0022$ \cite{Planck}.}
    \label{Mee1}
\end{figure}

\begin{figure}[H]
    \centering
    \includegraphics[width=1\linewidth]{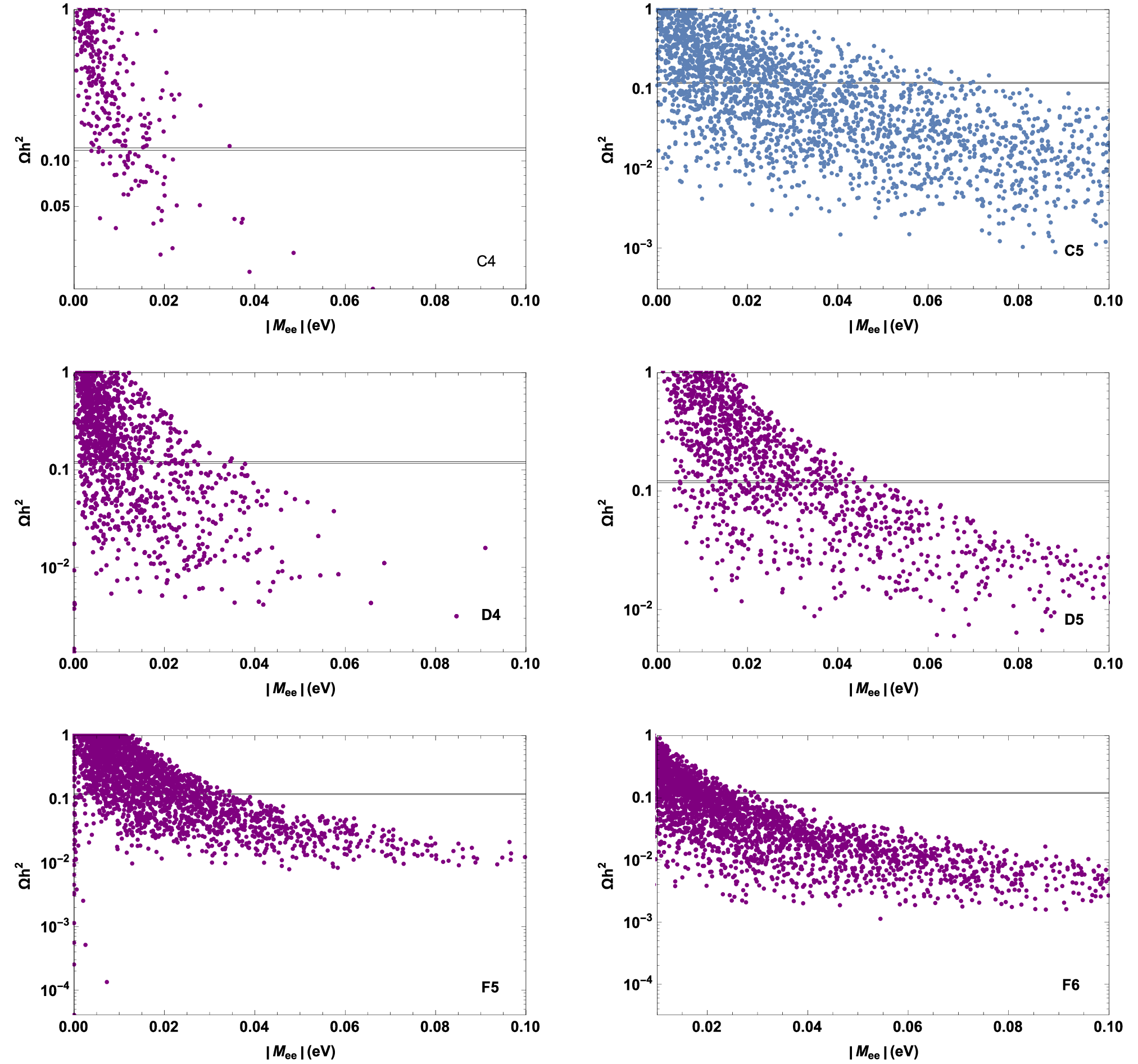}
    \caption{Correlation between Relic density of dark matter $(\Omega h^2)$ and effective Majorana mass $(|M_{ee}|)$ in neutrinoless double beta decay for textures $C_4$,$D_{4,5}$, $F_{5,6}$. The horizontal lines indicating the measured relic density value, $\Omega h^2 = 0.1191 \pm 0.0022$ \cite{Planck}.}
    \label{Mee2}
\end{figure}

In Section 3, we explain the relationship between each component of the neutrino mass matrix and the effective Majorana mass pertinent to neutrinoless double beta decay. We presented a total of twelve configurations of the neutrino mass matrix, each containing one zero element and one null sub-trace. We have obtained the theoretical framework necessary for calculating the relic density of dark matter and its relationship to the effective Majorana mass in neutrinoless double beta decay.\\
The equations that form the basis of our analysis are presented in Table $\ref{Tab2}$ and Table $\ref{Tab22}$ in the Appendix. These formulas are utilised to ascertain the Yukawa couplings $h_{\alpha\beta}$, which subsequently act as critical inputs in the computation of the dark matter relic density. This analysis examines the normal hierarchy of the neutrino mass matrix, defined by the mass ordering $M_1 \leq M_2 < M_3$.  Assuming a normal mass hierarchy, the \( f^X_{\alpha\beta} \) coefficients are calculated by randomly varying the Dirac CP-violating phase \( \delta \) within the ranges outlined for various texture structures, as indicated in Table~\ref{tab303}. Simultaneously, the mixing angles (\(\theta_{12}\), \(\theta_{23}\), \(\theta_{13}\)) are randomly varied within their corresponding \(3\sigma\) experimental ranges provided in Table \ref{tab33}.\\
We randomly vary the mass \( M_1 \) while maintaining the hierarchy \( M_1 \leq M_2 < M_3 \) and \( M_0 \geq M_1 \), along with the quartic coupling \( \lambda \), diagonal Yukawa couplings \( (h_{e1}, h_{\mu2}, h_{\tau3}) \), and effective Majorana mass \( |M_{ee}| \) within the specified ranges outlined in Table \ref{tab3}. Substituting these values into the equations provided for each texture in Tables \ref{Tab2} and \ref{Tab22}, and solving them concurrently for the off-diagonal couplings ($h_{e2}, h_{e3}, h_{\mu1}, h_{\mu3}, h_{\tau1}, h_{\tau2}$) with a constraint of ($0-1.5$) on their values. Upon computing the off-diagonal coupling, we utilize them to determine the thermal average cross section as presented in Eqn.(\ref{2.15}) through Eqn.(\ref{2.14}). The relic density of dark matter was then computed using Eqn.(\ref{2.16}). Furthermore, the limitation on the branching ratio from lepton flavor-violating processes, $\mu \rightarrow e \gamma$ with $Br(\mu \rightarrow e \gamma) \leq 4.2 \times 10^{-13}$, was integrated into our study \cite{MEG:2016leq}.\\
Fig.\ref{M1} and $\ref{M2}$ depict the correlation between the relic density of dark matter($\Omega h^2$), and the dark matter mass($M_1$) for the given textures, with the horizontal lines indicating the measured relic density value, $\Omega h^2 = 0.1191 \pm 0.0022$ \cite{Planck}. This measured value constrain the mass of dark matter. Likewise, Fig.$\ref{Mee1}$ and Fig.$\ref{Mee2}$ illustrate the correlation between the effective Majorana mass, $|M_{ee}|$, and the dark matter mass, $M_1$, with the horizontal line representing the measured value of dark matter relic density. The measured relic density establishes a lower limit on $|M_{ee}|$.\\
The plots in Fig.$\ref{M1}$ and $\ref{M2}$ indicate a particular range of dark matter mass ($M_1$) that aligns with the observed relic density of dark matter. It provides an upper limit on the dark matter mass for each  texture.\\
It is worth noting that all the correlation plots provide an upper limit on the DM mass for the Yukawa coupling ranges of ($0-1.5$). Nevertheless, the observed relic density of DM could not be satisfied by texture $C_5$ (shown in different colors in respective plots). In addition, this texture satisfies the relic density of DM when the full allowed range of Yukawa coupling is taken into account, which is $h_{\alpha\beta}\leq\sqrt{4\pi}$ \cite{Allwicher:2021rtd}. However, texture $C_5$ could not provide any bound on dark matter mass and effective Majorana mass in neutrinoless double beta decay.  However, this is not the exclusive approach to validate or refute the textures, as we can ascertain them by observation in both current and future research.  The experiments, such as SuperNEMO, KamLAND-Zen, NEXT, and nEXO (5 years), exhibit sensitivity thresholds of 0.05 eV, 0.045 eV, 0.03 eV, and 0.015 eV, respectively \cite{Barabash:2011row,KamLAND-Zen:2016pfg,NEXT:2013wsz,NEXT:2009vsd,Licciardi:2017oqg}.
  
  \begin{table}[H]
  \centering
  \begin{tabular}{|P{2cm}|P{3cm}|P{3cm}|}
    \hline
    \textbf{Texture} & \textbf{$M_{1}$ (TeV)} & \textbf{$\left|M_{ee}\right|$ (eV)} \\
    \hline
    $B_{1}$ & $\leq 2.52$ & $\geq 0.0046$ \\
    \hline
    $B_{4}$ & $\leq 5.45$ & $\geq 0.0039$ \\
    \hline
    $B_{5}$ & $\leq 2.45$ & $\geq 0.0067$ \\
    \hline
    $C_{1}$ & $\leq 1.82$ & $\geq 0.0036$ \\
    \hline
    $C_{2}$ & $\leq 1.54$ & $\geq 0.0039$ \\
    \hline
    $C_{3}$ & $\leq 4.58$ & $\geq 0.0096$ \\
    \hline
    $C_{4}$ & $\leq 2.31$ & $\geq 0.0039$ \\
    \hline
    $C_{5}$ & \multicolumn{2}{|l|}{disallowed by the large value of Yukawa couplings} \tabularnewline 
    \hline
    $D_{4}$ & $\leq 6.41$ & $[0-0.1]$ \\
    \hline
    $D_{5}$ & $\leq 2.81$ & $\geq 0.0036$ \\
    \hline
    $F_{5}$ & $\leq 2.22$ & $[0-0.1]$ \\
    \hline
    $F_{6}$ & $\leq 7.36$ & $[0-0.1]$ \\
    \hline
  \end{tabular}
  \caption{Bounds on DM mass $M_1$ and $|M_{ee}|$ for allowed  textures.}
  \label{T3}
\end{table}

\section{Conclusions}
This work focuses on the Scotogenic model and effective Majorana mass in $0\nu\beta\beta$ decay. We carry out our work on the texture structures of the neutrino mass matrices that contained one zero element and one vanishing sub-trace. Here we have considered only twelve textures, $B_{1,4,5}$, $C_{1,2,...5}$, $D_{4,5}$, $F_{5,6}$ which are allowed only for NH under the $3\sigma$ neutrino oscillation data. In this work we have established mathematical correlations between Dark matter and $|M_{ee}|$ for each of the twelve considered textures. Our calculations account for the DM relic density, the DM mass $M_1$, and $|M_{ee}|$.\par 
Now we summarize our study as follows:
\begin{itemize}
    \item[1.]In Fig.\ref{M1} and Fig. \ref{M2} we observe the predicted parameter space for dark matter mass ($M_1$) and relic density of dark matter ($\Omega h^2$) for textures ($B_{1,4,5}$, $C_{1,2,...5}$) and ($D_{4,5}$, $F_{5,6}$). The horizontal line in the figure shows the measured value of the dark matter relic density ($\Omega h^2=0.120 \pm 0.001$), which serves to constrain $M_1$.
    \item[2.] Again, in Fig.\ref{Mee1} and Fig.\ref{Mee2} we observe the predicted parameter space for effective Majorana mass ($|M_{ee}|$) in neutrinoless double beta decay and relic density of dark matter ($\Omega h^2$). Here also the horizontal line is the measured value of the DM relic density, which serves to constrain $|M_{ee}|$.
    \item[3.] All of these plots are for the [0,1.5] range of Yukawa couplings, with the exception of $C_5$, which does not meet the DM relic density requirements for this range. The condition on the DM relic density for coupling $\leq\sqrt{4\pi}$ is satisfied, nevertheless. This large amount, however, does not bode well.
    \item[4.] We also observe that the DM mass falls somewhere between 1.54 and 7.36 TeV from Table \ref{T3}. We also obtained a lower bound on $|M_{ee}|$, with the exception of $C_5$, $D_4$, $F_5$, and $F_6$.
\end{itemize}
   We expect that the predicted values of the parameters will be verified in the current and the future experiments designed for the purpose.

\section{Appendix }
\begin{align}
\left.
 \begin{array}{lll}
	f_{e\tau}^{B_1}=\frac{I_1 \sin{2\theta_{12}}(I_1 c_{13}^2-1)\tan{2\theta_{13}}}{-e^{i \delta}R_4c_{23}\sin{2\theta_{12}}+4e^{4i \delta}\cos{2\theta_{12}}s_{13}s_{23}}
\\
f_{\mu\mu}^{B_1}=\frac{c_{23}\sin{2\theta_{12}}(-e^{2i\delta}(-1+I_2)+4s_{13}^2s_{23}^2)+2e^{i\delta}\cos{2\theta_{12}}s_{13}(s_{23}-\sin{3\theta_{23}})}{-2e^{2i\delta}\cos{2\theta_{23}}c_{23}\sin{2\theta_{{12}}}+4e^{3i\delta}s_{13}I_3}
\\
f_{\mu\tau}^{B_1}=\frac{s_{23}(\sin{2\theta_{12}}(e^{2i\delta}(1+I_2)+2\cos{2\theta_{23}}\sin^2{\theta_{13}})+4e^{i\delta}\cos{2\theta_{12}}s_{13}\sin{2\theta_{23}})}{-2e^{2i\delta}\cos{2\theta_{13}}c_{23}\sin{2\theta_{12}}+4e^{3i\delta}s_{13}I_3}
\end{array}
\right\}
\end{align}
\begin{align}
\left.
 \begin{array}{lll}
f_{e\tau}^{B_4}=\frac{(-4e^{2i\delta}c_{23}^2+e^{4i\delta}J_1+\cos{2\theta_{23}})\sin{2\theta_{12}}\sin{2\theta_{13}}}{-16 e^{4i\delta}\cos{2\theta_{12}}s_{12}\sin^3{\theta_{23}}-4e^{3i\delta}c_{23}\sin{2\theta_{12}}J_2}
\\
f_{\mu\mu}^{B_4}=\frac{-4e^{4i\delta}c_{12}c^3_{23}s_{12}s^2_{13}-4e^{3i\delta}\cos{2\theta_{12}}c^2_{23}s_{12}s_{23}+c_{23}\sin{2\theta_{12}}(-e^{2i\delta}J_3+4s^2{13}s^2_{23})}{-e^{2i\delta}c_{23}(-J_2+2e^{2i\delta}s^2_{13}s^2_{23}+\cos{2\theta_{23}})\sin{2\theta_{12}}+e^{3i\delta}s_{13}s_{23}(4\cos{2\theta_{12}}s^2_{23}+e^{i\delta}\sin{2\theta_{12}}s_{13}\sin{2\theta_{23}})}
\\
f_{\mu\tau}^{B_4}=\frac{8s_{23}(-\sin{2\theta_{12}}(e^{2i\delta}(\cos{2\theta_{13}}+\cos{2\theta_{23}})+(1+e^{4i\delta})c^2_{23}s^2{13})-4e^{2i\delta}\cos{d}\cos{2\theta_{12}}c_{23}s_{13}s_{23})}{-16 e^{3i\delta}\cos{2\theta_{12}}s_{13}s^3_{23}-4e^{3i\delta}c_{23}\sin{2\theta_{12}}J_2}
\end{array}
\right\}
\end{align}
\begin{align}
\left.
 \begin{array}{lll}
f_{e\tau}^{B_5}=\frac{4c_{12}c_{13}\csc^2{\theta_{23}}(c_{23}+e^{2i\delta}(-c_{23}+\csc{\theta_{23}}))s_{12}s_{13}}{e^{i\delta}k_1\csc^2{\theta_{23}}\sin{2\theta_{12}}+4e^{2i\delta}\cos{2\theta_{12}}\cot{\theta_{23}}s_{13}+2e^{3i\delta}\cot^2{\theta_{23}}\sin{2\theta_{12}}s^2_{12}}
\\
f_{\mu\mu}^{B_5}=\frac{4s_{13}(e^{i\delta}\cos{2\theta_{12}}c_{23}(e^{2id}c_{23}+s_{23})+c_{12}s_{12}s_{13}C_2)+e^{2id}\sin{2\theta_{12}}(1+\cos{2\theta_{23}}-\sin{2\theta_{23}})}{e^{2i\delta}K_1+4e^{4i\delta}c_{12}\cos^2{\theta_{23}}s_{12}s^2{13}+2e^{3i\delta}\cos{2\theta_{12}}s_{13}\sin{2\theta_{23}}}
\\
f_{\mu\tau}^{B_5}=\frac{-e^{-2i\delta}(-32e^{i\delta}\cos{2\theta_{12}}s_{13}s_{23}(c_{23}+e^{2i\delta}s_{23})+\sin{2\theta_{23}}(-16c^2_{23}s^2{13}-8e^{2i\delta}C_3))}{8(k_1+2e^{i\delta}s_{13}(e^{i\delta}c^2{23}\sin{2\theta_{12}}s_{13}+\cos{2\theta_{12}}\sin{2\theta_{23}}))}
\end{array}
\right\}
\end{align}
\begin{align}
\left.
 \begin{array}{lll}
f_{e\mu}^{C_1}=\frac{(1+e^{2i\delta})\sin{2\theta_{12}}\sin{2\theta_{13}}}{-4e^{2i\delta}\cos{2\theta_{12}}c_{23}s_{12}+2e^{i\delta}\sin{2\theta_{12}}L_1s_{23}}
\\
f_{\mu\mu}^{C_1}=\frac{e^{i\delta}\cos{2\theta_{12}}(c_{23}+\cos{3\theta_{23}})s_{13}+\sin{2\theta_{12}}L_2s_{23}}{-2e^{3i\delta}\cos{2\theta_{12}}c_{23}s_{12}+e^{2i\delta}\sin{2\theta_{12}}L_1s_{23}}
\\
f_{\tau\tau}^{C_1}=\frac{c_{23}(\sin{2\theta_{12}}D_3-4e^{i\delta}\cos{2\theta_{12}}s_{13}\sin{2\theta_{23}})}{-4e^{3i\delta}\cos{2\theta_{12}}c_{23}s_{12}+2e^{2i\delta}\sin{2\theta_{12}}L_3s_{23}}
\end{array}
\right\}
\end{align}
\begin{align}
\left.
 \begin{array}{lll}
f_{e\mu}^{C_2}=\frac{(e^{2i\delta}+\cot^2{\theta_{23}})\csc{z}\sin{2x}\sin{2y}}{-2e^{2i\delta}\cos{2\theta_{23}}\sin{2\theta_{12}}+2e^{3id}\sin{2\theta_{12}}s_{13}^2-e^{2id}M_1}
\\
f_{\mu\mu}^{C_2}=\frac{-8e^{id}\cos{2\theta_{12}}(c_{23}+\cos{3\theta_{23}})s_{13}+4\sin{2\theta_{12}}(2c_{23}^2s_{13}(e^{3i\delta}c_{13}+2s_{13}s_{23})+e^{i\delta}s_{23}M_3}{16e^{3i\delta}\cos{2\theta_{12}}c_{23}s_{13}s_{23}^2+\sin{2\theta_{12}}M_2}
\\
f_{\mu\tau}^{C_2}=\frac{c_{23}(8c_{23}^2\sin{2\theta_{12}}s_{12}^2-8e^{2i\delta}\sin{2\theta_{12}}s_{23}(i \sin{\delta}\sin{2\theta_{12}}+s_{23})+8e^{i\delta}\cos{2\theta_{12}}s_{13}\sin{2\theta_{23}})}{16e^{3i\delta}\cos{2\theta_{12}}c_{23}s_{13}s_{23}^2+\sin{2\theta_{12}}M_2}
\end{array}
\right\}
\end{align}
\begin{align}
\left.
 \begin{array}{lll}
f_{e\mu}^{C_3}=\frac{(-1+e^{2i\delta})\sin{2x}\sin{2y}\sin{2z}}{2(e^{i\delta}c_{23}N_2+e^{2i\delta}N_1+e^{3i\delta}\sin{2\theta_{12}}s_{13}^2s_{23}\sin{2\theta_{23}})}
\\
f_{\mu\mu}^{C_3}=\frac{e^{2i\delta}c_{23}N_2+e^{3i\delta}c_{23}^2\sin{2\theta_{12}}\sin{2\theta_{13}}+e^{i\delta}s_{23}N_3+\sin{2\theta_{12}}s_{13}^2s_{23}\sin{2\theta_{23}}}{e^{2i\delta}c_{23}N_2+e^{3i\delta}N_1+e^{4i\delta}\sin{2\theta_{12}}s_{13}^2s_{23}\sin{2\theta_{23}}}
\\
f_{\mu\tau}^{C_3}=\frac{-2\sin{2\theta_{12}}c_{23}^2s_{13}^2+2e^{2i\delta}\sin{2\theta_{12}}(c_{23}+e^{i\delta}c_{13}s_{13})s_{23}^2+e^{i\delta}c_{23}^2(N_1-4\cos{2\theta_{12}}s_{13}s_{23}s_{23}^2)}{e^{2i\delta}c_{23}N_2+e^{3i\delta}N_1+e^{4i\delta}\sin{2\theta_{12}}s_{13}^2s_{23}\sin{2\theta_{23}}}
\end{array}
\right\}
\end{align}
\begin{align}
\left.
 \begin{array}{lll}
f_{e\mu}^{C_4}=\frac{-e^{-i\delta}(1+e^{2i\delta}(-3+\cos{2\theta_{23}})-\cos{2\theta_{23}})\csc^2{\theta_{23}}\sin{2\theta_{12}}\sin{2\theta_{13}}}{8(\cot{\theta_{23}}s_{12}+e^{i\delta}s_{13}c_{12})P_1}
\\
f_{\mu\mu}^{C_4}=\frac{e^{-2i\delta}(4e^{2i\delta}P_2\sin{2\theta_{12}}+16e^{i\delta}\cos{2\theta_{12}}(e^{2i\delta}c_{23}^2+\cos{2\theta_{23}})\cot{\theta_{23}}s_{13}-8(2+e^{4i\delta})c_{23}^2\sin{2\theta_{12}}s_{13}^2)}{16(c_{23}s_{12}+e^{i\delta}s_{13}c_{12}s_{23})P_1}
\\
f_{\mu\tau}^{C_4}=\frac{e^{-2i\delta}(\cot{\theta_{23}}\sin{2\theta_{12}}(e^{2i\delta}(1+\cos{2\theta_{13}}-2\cos{2\theta_{23}})+e^{i\delta}s_{13}P_3)}{4(c_{23}s_{12}+e^{i\delta}s_{13}c_{12}s_{23})P_1}
\end{array}
\right\}
\end{align}
\begin{align}
\left.
 \begin{array}{lll}
f_{e\mu}^{C_5}=\frac{2c_{12}c_{13}\sec{\theta_{23}}s_{12}s_{13}(e^{2i\delta}(\sec{\theta_{23}}-s_{23})+s_{23})}{-2e^{2i\delta}c_{12}s_{13}s_{23}+e^{i\delta}\sin{2\theta_{12}}Q_1}
\\
f_{\mu\mu}^{C_5}=\frac{s_{13}Q_4+e^{2i\delta}\sin{2\theta_{12}}Q_3}{e^{2i\delta}(\cos{2\theta_{13}}-\cos{2\theta_{23}})\sin{2\theta_{12}}+Q_2}
\\
f_{\tau\tau}^{C_5}=\frac{e^{-2i\delta}(e^{2i\delta}\sin{2\theta_{12}}(-1+\cos{2\theta_{23}}+\sin{2\theta_{23}})+4s_{13}Q_4)}{(-\cos{2\theta_{13}}+\cos{2\theta_{23}})\sin{2\theta_{12}}-Q_2}
\end{array}
\right\}
\end{align}
\begin{align}
\left.
 \begin{array}{lll}
f_{e\mu}^{D_4}=\frac{e^{-id}(2e^{i\delta}\cos{2\theta_{12}}(e^{2i\delta}c^2_{23}+\cos{2\theta_{23}})\sin{2\theta_{13}}s_{23}+c_{13}c_{23}R_1)}{2(-2e^{i\delta}\cos{2\theta_{12}}\cos{2\theta_{13}}\cos{2\theta_{13}}+\sin{2\theta_{12}}s_{13}R_2)}
\\
f_{e\tau}^{D_4}=\frac{-2e^{i\delta}\cos{2\theta_{12}}c_{23}(e^{2i\delta}c^2{23}+\cos{2\theta_{23}})\sin{2\theta_{13}}+c_{13}R_3s_{23}}{4e^{2i\delta}\cos{2\theta_{12}}\cos{2\theta_{13}}\cos{2\theta_{23}}-2e^{i\delta}\sin{2\theta_{12}}s_{13}R_2\sin{2\theta_{23}}}
\\
f_{\mu\tau}^{D_4}=\frac{e^{i\delta}\sin{2\theta_{12}}s_{13}(1-\cos{2\theta_{13}}(-2+\cos{2\theta_{23}})+2e^{2i\delta}c^2_{23}s^2_{23})-2e^{2i\delta}\cos{2\theta_{12}}\cos{2\theta_{13}}\sin{2\theta_{23}}}{4e^{2i\delta}\cos{2\theta_{12}}\cos{2\theta_{13}}\cos{2\theta_{23}}-R_4\sin{2\theta_{12}}s_{13}\sin{2\theta_{23}}}
\end{array}
\right\}
\end{align}
\begin{align}
\left.
 \begin{array}{lll}
f_{e\mu}^{D_5}=\frac{e^{-i\delta}(2e^{i\delta}\cos{2\theta_{12}}\sin{2\theta_{13}}(e^{2i\delta}c_{23}+s_{23})\sin{2\theta_{23}}s_{23}+c_{13}S_1)}{4(e^{2i\delta}c^2_{23}\sin{2\theta_{12}})}
\\
f_{e\tau}^{D_5}=\frac{e^{-i\delta}c_{13}(-8c_{23}s_{13}G_3)-2e^{2i\delta}\sin{2\theta_{13}}(-c_{23}-\cos{3\theta_{23}}+2\cos{2\theta_{13}}(c_{23}-2s_{23})+s_{23}+\sin{3\theta_{23}})}{8(e^{2i\delta}c^2_{23}\sin{2\theta_{12}}s^3_{13}+\cos{2\theta_{13}}S_2)}
\\
f_{\tau\tau}^{D_5}=\frac{2e^{i\delta}\cos{2\theta_{12}}\cos{2\theta_{13}}\cos{2\theta_{23}}+\sin{2\theta_{12}}s_{13}S_4}{e^{2i\delta}c^2{23}\sin{2\theta_{12}}s^3_{13}+\cos{2\theta_{13}}S_2}
\end{array}
\right\}
\end{align}
\begin{align}
\left.
 \begin{array}{lll}
f_{e\mu}^{F_5}=\frac{-(e^{i\delta}c_{13}(2s_{13}^2\sin{2\theta_{12}}\cos{\theta_{23}}(c_{23}+e^{4i\delta}s_{23})+8e^{i\delta}\cos{2\theta_{12}}s_{13}s_{23}^2W_1))}{(4(e^{2i\delta}\sin{2\theta_{12}}s_{23}^2s_{12}^3+\cos{2\theta_{12}}W_3))}
\\
f_{e\tau}^{F_5}=\frac{e^{-i\delta}(-2e^{i\delta}\cos{2\theta_{12}}\sin{2\theta_{13}}\sin{2\theta_{23}}W_1+c_{13}(-4\cos{2\theta_{12}}s_{13}^2(c_{23}^3-e^{4i\delta}s_{23}^3)-e^{2i\delta}\sin{2\theta_{12}}(s_{23}-c_{23}+\cos{3\theta_{23}}+\sin{3\theta_{23}})))}{(4(e^{2i\delta}\sin{2\theta_{12}}s_{23}^2s_{12}^3+\cos{2\theta_{12}}W_3))}
\\
f_{\tau\tau}^{F_5}=\frac{2e^{i\delta}\cos{2\theta_{12}}\cos{2\theta_{13}}\cos{2\theta_{23}}+\sin{2\theta_{12}}s_{13}W_4}{e^{2i\delta}\sin{2\theta_{12}}s_{13}^3s_{23}^2+\cos{2\theta_{13}}W_3}
\end{array}
\right\}
\end{align}
\begin{align}
\left.
 \begin{array}{lll}
f_{e\mu}^{F_6}=\frac{e^{-i\delta}c_{13}c_{23}\sin{2\theta_{12}}(2-3\cos{2\theta_{13}}-3\cos{2\theta_{23}}+2e^{2i\delta}s^2_{13}s^2_{12})+\sin{2\theta_{13}}s_{23}V_3}{V_1+2e^{i\delta}\sin{2\theta_{12}}s_{13}V_2\sin{2z}}
\\
f_{e\tau}^{F_6}=\frac{c_{13}(s_{23}J_4+2e^{i\delta}\cos{2\theta_{12}}s_{13}(-c_{23}-\cos{3\theta_{23}}+e^{2i\delta}\sin{2\theta_{23}}s_{23}))}{V_1+2e^{i\delta}\sin{2\theta_{12}}s_{13}V_2\sin{2\theta_{23}}}
\\
f_{\tau\tau}^{F_6}=\frac{\sin{2\theta_{12}}s_{13}J_5+2e^{i\delta}\cos{2\theta_{12}}\cos{2\theta_{12}}\sin{2\theta_{23}}}{V_1+2\sin{2\theta_{12}}s_{13}V_2\sin{2\theta_{23}}}
\end{array}
\right\},
\end{align}
where $c_{ij}\equiv\cos\theta_{ij}$ and $s_{ij}\equiv\sin\theta_{ij}$.\\

\newpage
\begin{align}
\left.
 \begin{array}{lllllllllllllllllllllllllllllllllllll}
 I_1=1+e^{2i\delta},\\ I_2=\cos{2\theta_{12}}+2\cos{2\theta_{23}},\\ I_3=e^{i\delta}c_{12}c_{23}s_{12}s_{13}+\cos{2\theta_{12}}s_{23},\\
J_1=-6+\cos{2\theta_{23}},\\ J_2=1-2\cos{2\theta_{13}}-\cos{2\theta_{23}}+2e^{2i\delta}s^2_{12}s^2_{23},\\ J_3=-2+\cos{2\theta_{13}}+3\cos{2\theta_{23}},\\
 K_1=(\cos{2\theta_{13}}+\cos{2\theta_{23}})\sin{2\theta_{12}}, \\K_2=e^{4i\delta}c^2_{23}\cot{\theta_{23}}-s^2_{23}, \\K_3=\cos{2\theta_{23}}+\cos{2\theta_{13}}(1-2\cot{\theta_{23}}+(1+e^{2i\delta}s^2{13})\sin{2\theta_{23}}),\\
 L_1=-\cos{2\theta_{13}}+e^{2i\delta}s^2_{12}, \\L_2=-2c^2_{23}s^2{12}-e^{2i\delta}(\cos{2\theta_{23}}+s^2{13}),\\ 
 L_3=e^{2i\delta}(1+\cos{2\theta_{13}}-2\cos{2\theta_{23}})-2\cos{2\theta_{23}}s^2_{12},\\
 M_1=4\cos{2\theta_{12}}\cot{\theta_{23}}s_{13}+\csc^3{\theta_{23}}\sin{2\theta_{12}}\sin{2\theta_{12}}, \\M_2=4e^{3i\delta}\sin{2\theta_{12}}+8e^{2i\delta}c_{23}^2s_{23}-8e^{4i\delta}s_{13}^2s_{23}^3,\\ 
 M_3=2e^{i\delta}(\cos{2\theta_{23}}+s_{13}^2)+\sin{2\theta_{12}}s_{23},\\
 N_1=\sin{2\theta_{12}}\sin{2\theta_{13}}-4\cos{2\theta_{12}}c_{23}^2s_{13}s_{23}, \\N_2=(\cos{2\theta_{13}}-\cos{2\theta_{23}})\sin{2\theta_{23}},\\ 
N_3=-4\cos{2x}c_{23}^2s_{13}+\sin{2\theta_{12}}\sin{2\theta_{13}}s_{23},\\
 P_1=-c_{12}c_{23}+e^{i\delta}s_{12}s_{13}s_{23}, \\P_2=-2+3\cos{2\theta_{13}}-3\cos{2\theta_{23}},\\ 
 P_3=-4(1+e^{2i\delta})\cos{2\theta_{12}}c_{23}^2+e^{3i\delta}\sin{2\theta_{12}}s_{13}\sin{2\theta_{23}},\\
 Q_1=-\sec{\theta_{23}}\sin(\theta_{12}-\theta_{13})\sin(\theta_{12}+\theta_{13})+e^{2i\delta}s^2{13}s_{23}t_{23}, \\Q_2=2e^{4i\delta}\sin{2\theta_{12}}s^2_{13}s^2{23}-2e^{3i\delta}\cos{2\theta_{12}}s_{13}\sin{2\theta_{23}},
 \\Q_3=-\cos{2z\theta_{23}}+\sin{2\theta_{23}}+\cos{2\theta_{13}}(1-2t_{23}),\\
 Q_4=(-4e^{3i\delta}\cos{2\theta_{12}}c^2_{23}+2\sin{2\theta_{12}}s_{13}s^2_{23}-2e^{i\delta}\cos{2\theta_{12}}\sin{2\theta_{23}}+e^{4i\delta}\sin{2\theta_{12}}s_{13}\sin{2\theta_{23}}),\\
R_1=e^{2i\delta}(-2+\cos{2\theta_{13}}+3\cos{2\theta_{23}})\sin{2\theta_{12}}+4c_{12}s_{12}s^2{13}(e^{4i\delta}c^2_{23}-2s^2_{23}),\\
R_2=(\cos{2\theta_{13}}-e^{2i\delta}s^2{\theta_{13}})\sin{2\theta_{23}}, \\
R_3=4e^{4i\delta}c_{12}\cos^2{z\theta_{23}}s_{12}s^2_{13}+\sin{2\theta_{13}}(e^{2i\delta}(2-3\cos{2\theta_{13}}+3\cos{2\theta_{23}})+4c^2{23}s^2{13}),\\
R_4=-1+(1+2e^{i\delta})\cos{2\theta_{13}},\\
S_1=8e^{4i\delta}c_{12}\cos^3{\theta_{23}}s_{12}s^2_{12}-4\sin{2\theta_{12}}s^2_{13}s^3_{23}+e^{2i\delta}\sin{2\theta_{12}}(-c_{23}+\cos{3\theta_{23}}+s_{23}+\sin{3\theta_{23}}), \\
S_2=\sin{2\theta_{23}}s_{13}s^2_{23}-e^{i\delta}\cos{2\theta_{12}}\sin{2\theta_{23}},\\
S_3=-2e^{i\delta}\cos{2\theta_{12}}c_{23}(e^{2i\delta}c_{23}+s_{23})+\sin{2\theta_{12}}s_{13}s_{23}(e^{4i\delta}c_{23}+s_{23}),\\
S_4=c^2_{13}-(\cos{2\theta_{13}}-e^{2i\delta}s^2_{13})\sin{2\theta_{23}},\\ W_1=c_{23}+e^{2i\delta}s_{23},\\
 W_2=(-1+4\cos{2\theta_{13}})c_{23}+\cos{3\theta_{23}}-2(\cos{2\theta_{13}}+\cos{2\theta_{23}})s_{23},\\
 W_3=c_{23}^2\sin{2\theta_{12}}s_{13}+e^{i\delta}\cos{2\theta_{12}}\sin{2\theta_{23}},\\
 W_4=c_{13}^2-(\cos{2\theta_{13}}-e^{2i\delta}s_{13}^2)\sin{2\theta_{23}} \\ V_1=4e^{2id}\cos{2\theta_{12}}\cos{2\theta_{13}}\cos{2\theta_{23}},\\
 V_2=-\cos{2\theta_{13}}+e^{2id}s^2_{13},\\
 V_3=(2e^{i\delta}\cos{2\theta_{12}}(-\cos{2\theta_{23}}+e^{2i\delta}s^2_{23})+\sin{2\theta_{12}}s_{13}\sin{2\theta_{23}}),\\
 V_4=-e^{2i\delta}(-2+\cos{2\theta_{13}}-3\cos{2\theta_{23}})\sin{2\theta_{12}}+2\sin{2\theta_{12}}s^2_{13}(2c^2_{23}-e^{4i\delta}s^2{23}),\\
 V_5=1+\cos{2\theta_{13}}(2+\cos{2\theta_{23}}+2e^{2i\delta}s^2_{13}s^2_{23}).
\end{array}
\right\}
\end{align}

\begin{table}[H]
    \centering
    \begin{tabular}{|c|c|}
    \hline
      Texture   &  Constraining Equations \\
      \hline
        $B_1$      &  $\begin{array} {lcl}h_{e1}^{2}\Lambda_{1}+h_{e2}^{2}\Lambda_{2}+h_{e3}^{2}\Lambda_{3}&=& M_{ee}\\
h_{e1}h_{\mu1}\Lambda_{1}+h_{e2}h_{\mu2}\Lambda_{2}+h_{e3}h_{\mu3}\Lambda_{3}&=& 0\\
h_{e1}h_{\tau1}\Lambda_{1}+h_{e2}h_{\tau2}\Lambda_{2}+h_{e3}h_{\tau3}\Lambda_{3}&=&f_{e\tau}^{B_1} M_{ee}\\
h_{\mu1}^{2}\Lambda_{1}+h_{\mu2}^{2}\Lambda_{2}+h_{\mu3}^{2}\Lambda_{3}&=&f_{\mu\mu}^{B_1} M_{ee}\\
h_{\mu1}h_{\tau1}\Lambda_{1}+h_{\mu2}h_{\tau2}\Lambda_{2}+h_{\mu3}h_{\tau3}\Lambda_{3}&=&f_{\mu\tau}^{B_1} M_{ee}\\
h_{\tau1}^{2}\Lambda_{1}+h_{\tau2}^{2}\Lambda_{2}+h_{\tau3}^{2}\Lambda_{3}&=& -f_{\mu\mu}^{B_1} M_{ee} \end{array}$  \\
         \hline
               
  $B_4$     &     $\begin{array} {lcl}h_{e1}^{2}\Lambda_{1}+h_{e2}^{2}\Lambda_{2}+h_{e3}^{2}\Lambda_{3}&=& M_{ee}\\
h_{e1}h_{\mu1}\Lambda_{1}+h_{e2}h_{\mu2}\Lambda_{2}+h_{e3}h_{\mu3}\Lambda_{3}&=&0\\
h_{e1}h_{\tau1}\Lambda_{1}+h_{e2}h_{\tau2}\Lambda_{2}+h_{e3}h_{\tau3}\Lambda_{3}&=&f_{e\tau}^{B_4}M_{ee}\\
h_{\mu1}^{2}\Lambda_{1}+h_{\mu2}^{2}\Lambda_{2}+h_{\mu3}^{2}\Lambda_{3}&=&f_{\mu\mu}^{B_4} M_{ee}\\
h_{\mu1}h_{\tau1}\Lambda_{1}+h_{\mu2}h_{\tau2}\Lambda_{2}+h_{\mu3}h_{\tau3}\Lambda_{3}&=&f_{\mu\tau}^{B_4} M_{ee}\\
h_{\tau1}^{2}\Lambda_{1}+h_{\tau2}^{2}\Lambda_{2}+h_{\tau3}^{2}\Lambda_{3}&=& -M_{ee}\end{array}$  \\
\hline
       $B_5$   &  $\begin{array} {lcl}h_{e1}^{2}\Lambda_{1}+h_{e2}^{2}\Lambda_{2}+h_{e3}^{2}\Lambda_{3}&=& M_{ee}\\
h_{e1}h_{\mu1}\Lambda_{1}+h_{e2}h_{\mu2}\Lambda_{2}+h_{e3}h_{\mu3}\Lambda_{3}&=& 0 \\
h_{e1}h_{\tau1}\Lambda_{1}+h_{e2}h_{\tau2}\Lambda_{2}+h_{e3}h_{\tau3}\Lambda_{3}&=&f_{e\tau}^{B_5} M_{ee}\\
h_{\mu1}^{2}\Lambda_{1}+h_{\mu2}^{2}\Lambda_{2}+h_{\mu3}^{2}\Lambda_{3}&=&f_{\mu\mu}^{B_5} M_{ee}\\
h_{\mu1}h_{\tau1}\Lambda_{1}+h_{\mu2}h_{\tau2}\Lambda_{2}+h_{\mu3}h_{\tau3}\Lambda_{3}&=&- M_{ee}\\
h_{\tau1}^{2}\Lambda_{1}+h_{\tau2}^{2}\Lambda_{2}+h_{\tau3}^{2}\Lambda_{3}&=& f_{\tau\tau}^{B_5} M_{ee}  \end{array}$ \\
      \hline
    $C_1$      & $\begin{array} {lcl}h_{e1}^{2}\Lambda_{1}+h_{e2}^{2}\Lambda_{2}+h_{e3}^{2}\Lambda_{3}&=& M_{ee}\\
h_{e1}h_{\mu1}\Lambda_{1}+h_{e2}h_{\mu2}\Lambda_{2}+h_{e3}h_{\mu3}\Lambda_{3}&=& f_{e\mu}^{C_1} M_{ee}\\
h_{e1}h_{\tau1}\Lambda_{1}+h_{e2}h_{\tau2}\Lambda_{2}+h_{e3}h_{\tau3}\Lambda_{3}&=&0\\
h_{\mu1}^{2}\Lambda_{1}+h_{\mu2}^{2}\Lambda_{2}+h_{\mu3}^{2}\Lambda_{3}&=&f_{\mu\mu}^{C_1} M_{ee}\\
h_{\mu1}h_{\tau1}\Lambda_{1}+h_{\mu2}h_{\tau2}\Lambda_{2}+h_{\mu3}h_{\tau3}\Lambda_{3}&=&f_{\mu\tau}^{C_1} M_{ee}\\
h_{\tau1}^{2}\Lambda_{1}+h_{\tau2}^{2}\Lambda_{2}+h_{\tau3}^{2}\Lambda_{3}&=& -f_{\mu\mu}^{C_1} M_{ee} \end{array} $  \\
         \hline
              
    $C_2$     &   $\begin{array} {lcl}h_{e1}^{2}\Lambda_{1}+h_{e2}^{2}\Lambda_{2}+h_{e3}^{2}\Lambda_{3}&=& M_{ee}\\
 h_{e1}h_{\mu1}\Lambda_{1}+h_{e2}h_{\mu2}\Lambda_{2}+h_{e3}h_{\mu3}\Lambda_{3}&=& f_{e\mu}^{C_2} M_{ee}\\
h_{e1}h_{\tau1}\Lambda_{1}+h_{e2}h_{\tau2}\Lambda_{2}+h_{e3}h_{\tau3}\Lambda_{3}&=&0\\
h_{\mu1}^{2}\Lambda_{1}+h_{\mu2}^{2}\Lambda_{2}+h_{\mu3}^{2}\Lambda_{3}&=&f_{\mu\mu}^{C_2} M_{ee}\\
h_{\mu1}h_{\tau1}\Lambda_{1}+h_{\mu2}h_{\tau2}\Lambda_{2}+h_{\mu3}h_{\tau3}\Lambda_{3}&=&f_{\mu\tau}^{C_2} M_{ee}\\
h_{\tau1}^{2}\Lambda_{1}+h_{\tau2}^{2}\Lambda_{2}+h_{\tau3}^{2}\Lambda_{3}&=& -f_{e\mu}^{C_2} M_{ee}  \end{array}$ \\
         \hline
               
   $C_3$      &  $\begin{array} {lcl}h_{e1}^{2}\Lambda_{1}+h_{e2}^{2}\Lambda_{2}+h_{e3}^{2}\Lambda_{3}&=& M_{ee}\\
h_{e1}h_{\mu1}\Lambda_{1}+h_{e2}h_{\mu2}\Lambda_{2}+h_{e3}h_{\mu3}\Lambda_{3}&=& f_{e\mu}^{C_3} M_{ee}\\
h_{e1}h_{\tau1}\Lambda_{1}+h_{e2}h_{\tau2}\Lambda_{2}+h_{e3}h_{\tau3}\Lambda_{3}&=&0\\
h_{\mu1}^{2}\Lambda_{1}+h_{\mu2}^{2}\Lambda_{2}+h_{\mu3}^{2}\Lambda_{3}&=&f_{\mu\mu}^{C_3} M_{ee}\\
h_{\mu1}h_{\tau1}\Lambda_{1}+h_{\mu2}h_{\tau2}\Lambda_{2}+h_{\mu3}h_{\tau3}\Lambda_{3}&=&-f_{e\mu}^{C_3} M_{ee}\\
h_{\tau1}^{2}\Lambda_{1}+h_{\tau2}^{2}\Lambda_{2}+h_{\tau3}^{2}\Lambda_{3}&=&f_{\tau\tau}^{T_{4}} M_{ee} \end{array} $  \\
         \hline

    \end{tabular}
\caption{Constraining equations relating loop factors and Yukawa couplings to the effective Majorana neutrino mass $\left|M_{ee}\right|$ for all the one zero textures with vanishing sub-trace.}
\label{Tab2}
\end{table}
\begin{table}[H]
    \centering
    \begin{tabular}{|c|c|}
    \hline
      Texture   &  Constraining Equations \\
      \hline
        $C_4$      &  $\begin{array} {lcl}h_{e1}^{2}\Lambda_{1}+h_{e2}^{2}\Lambda_{2}+h_{e3}^{2}\Lambda_{3}&=& M_{ee}\\
h_{e1}h_{\mu1}\Lambda_{1}+h_{e2}h_{\mu2}\Lambda_{2}+h_{e3}h_{\mu3}\Lambda_{3}&=& f_{e\mu}^{C_4} M_{ee}\\
h_{e1}h_{\tau1}\Lambda_{1}+h_{e2}h_{\tau2}\Lambda_{2}+h_{e3}h_{\tau3}\Lambda_{3}&=&0\\
h_{\mu1}^{2}\Lambda_{1}+h_{\mu2}^{2}\Lambda_{2}+h_{\mu3}^{2}\Lambda_{3}&=&f_{\mu\mu}^{C_4} M_{ee}\\
h_{\mu1}h_{\tau1}\Lambda_{1}+h_{\mu2}h_{\tau2}\Lambda_{2}+h_{\mu3}h_{\tau3}\Lambda_{3}&=&f_{\mu\tau}^{C_4} M_{ee}\\
h_{\tau1}^{2}\Lambda_{1}+h_{\tau2}^{2}\Lambda_{2}+h_{\tau3}^{2}\Lambda_{3}&=& - M_{ee} \end{array}$  \\
         \hline
               
  $C_5$     &     $\begin{array} {lcl}h_{e1}^{2}\Lambda_{1}+h_{e2}^{2}\Lambda_{2}+h_{e3}^{2}\Lambda_{3}&=& M_{ee}\\
h_{e1}h_{\mu1}\Lambda_{1}+h_{e2}h_{\mu2}\Lambda_{2}+h_{e3}h_{\mu3}\Lambda_{3}&=&f_{e\mu}^{C_4} M_{ee}\\
h_{e1}h_{\tau1}\Lambda_{1}+h_{e2}h_{\tau2}\Lambda_{2}+h_{e3}h_{\tau3}\Lambda_{3}&=&0\\
h_{\mu1}^{2}\Lambda_{1}+h_{\mu2}^{2}\Lambda_{2}+h_{\mu3}^{2}\Lambda_{3}&=&f_{\mu\mu}^{C_5} M_{ee}\\
h_{\mu1}h_{\tau1}\Lambda_{1}+h_{\mu2}h_{\tau2}\Lambda_{2}+h_{\mu3}h_{\tau3}\Lambda_{3}&=&- M_{ee}\\
h_{\tau1}^{2}\Lambda_{1}+h_{\tau2}^{2}\Lambda_{2}+h_{\tau3}^{2}\Lambda_{3}&=& f_{\tau\tau}M_{ee}\end{array}$  \\
\hline
       $D_4$   &  $\begin{array} {lcl}h_{e1}^{2}\Lambda_{1}+h_{e2}^{2}\Lambda_{2}+h_{e3}^{2}\Lambda_{3}&=& M_{ee}\\
h_{e1}h_{\mu1}\Lambda_{1}+h_{e2}h_{\mu2}\Lambda_{2}+h_{e3}h_{\mu3}\Lambda_{3}&=&f_{e\mu}^{D_5}M_{ee} \\
h_{e1}h_{\tau1}\Lambda_{1}+h_{e2}h_{\tau2}\Lambda_{2}+h_{e3}h_{\tau3}\Lambda_{3}&=&f_{e\tau}^{D_4} M_{ee}\\
h_{\mu1}^{2}\Lambda_{1}+h_{\mu2}^{2}\Lambda_{2}+h_{\mu3}^{2}\Lambda_{3}&=&0\\
h_{\mu1}h_{\tau1}\Lambda_{1}+h_{\mu2}h_{\tau2}\Lambda_{2}+h_{\mu3}h_{\tau3}\Lambda_{3}&=&f_{\mu\tau}^{D_5} M_{ee}\\
h_{\tau1}^{2}\Lambda_{1}+h_{\tau2}^{2}\Lambda_{2}+h_{\tau3}^{2}\Lambda_{3}&=& - M_{ee}  \end{array}$ \\
      \hline
    $D_5$      & $\begin{array} {lcl}h_{e1}^{2}\Lambda_{1}+h_{e2}^{2}\Lambda_{2}+h_{e3}^{2}\Lambda_{3}&=& M_{ee}\\
h_{e1}h_{\mu1}\Lambda_{1}+h_{e2}h_{\mu2}\Lambda_{2}+h_{e3}h_{\mu3}\Lambda_{3}&=& f_{e\mu}^{D_5} M_{ee}\\
h_{e1}h_{\tau1}\Lambda_{1}+h_{e2}h_{\tau2}\Lambda_{2}+h_{e3}h_{\tau3}\Lambda_{3}&=&f_{e\tau}^{D_5} M_{ee}\\
h_{\mu1}^{2}\Lambda_{1}+h_{\mu2}^{2}\Lambda_{2}+h_{\mu3}^{2}\Lambda_{3}&=&0\\
h_{\mu1}h_{\tau1}\Lambda_{1}+h_{\mu2}h_{\tau2}\Lambda_{2}+h_{\mu3}h_{\tau3}\Lambda_{3}&=&- M_{ee}\\
h_{\tau1}^{2}\Lambda_{1}+h_{\tau2}^{2}\Lambda_{2}+h_{\tau3}^{2}\Lambda_{3}&=& -f_{\tau\tau}^{D_5} M_{ee} \end{array} $  \\
         \hline
              
    $F_5$     &   $\begin{array} {lcl}h_{e1}^{2}\Lambda_{1}+h_{e2}^{2}\Lambda_{2}+h_{e3}^{2}\Lambda_{3}&=& M_{ee}\\
 h_{e1}h_{\mu1}\Lambda_{1}+h_{e2}h_{\mu2}\Lambda_{2}+h_{e3}h_{\mu3}\Lambda_{3}&=& f_{e\mu}^{F_5} M_{ee}\\
h_{e1}h_{\tau1}\Lambda_{1}+h_{e2}h_{\tau2}\Lambda_{2}+h_{e3}h_{\tau3}\Lambda_{3}&=&f_{e\tau}^{F_5} M_{ee}\\
h_{\mu1}^{2}\Lambda_{1}+h_{\mu2}^{2}\Lambda_{2}+h_{\mu3}^{2}\Lambda_{3}&=&f_{\mu\mu}^{F_5} M_{ee}\\
h_{\mu1}h_{\tau1}\Lambda_{1}+h_{\mu2}h_{\tau2}\Lambda_{2}+h_{\mu3}h_{\tau3}\Lambda_{3}&=&- M_{ee}\\
h_{\tau1}^{2}\Lambda_{1}+h_{\tau2}^{2}\Lambda_{2}+h_{\tau3}^{2}\Lambda_{3}&=& 0  \end{array}$ \\
         \hline
               
   $F_6$      &  $\begin{array} {lcl}h_{e1}^{2}\Lambda_{1}+h_{e2}^{2}\Lambda_{2}+h_{e3}^{2}\Lambda_{3}&=& M_{ee}\\
h_{e1}h_{\mu1}\Lambda_{1}+h_{e2}h_{\mu2}\Lambda_{2}+h_{e3}h_{\mu3}\Lambda_{3}&=& f_{e\mu}^{F_6} M_{ee}\\
h_{e1}h_{\tau1}\Lambda_{1}+h_{e2}h_{\tau2}\Lambda_{2}+h_{e3}h_{\tau3}\Lambda_{3}&=&f_{e\tau}^{F_6} M_{ee}\\
h_{\mu1}^{2}\Lambda_{1}+h_{\mu2}^{2}\Lambda_{2}+h_{\mu3}^{2}\Lambda_{3}&=&-M_{ee}\\
h_{\mu1}h_{\tau1}\Lambda_{1}+h_{\mu2}h_{\tau2}\Lambda_{2}+h_{\mu3}h_{\tau3}\Lambda_{3}&=&f_{\mu\tau}^{F_6} M_{ee}\\
h_{\tau1}^{2}\Lambda_{1}+h_{\tau2}^{2}\Lambda_{2}+h_{\tau3}^{2}\Lambda_{3}&=&0 \end{array} $  \\
         \hline

    \end{tabular}
\caption{Constraining equations relating loop factors and Yukawa couplings to the effective Majorana neutrino mass $\left|M_{ee}\right|$ for all the one zero textures with vanishing sub-trace.}
\label{Tab22}
\end{table}
\section*{Acknowledgements}
 \noindent  B. C. Chauhan is thankful to the Inter University Centre for Astronomy and Astrophysics (IUCAA) for providing necessary facilities during the completion of this work.

\end{document}